\documentclass[aps,pra,showpacs,twocolumn,floatfix]{revtex4}

\usepackage{graphicx} % Include figure files
\usepackage{amsmath}
\usepackage{bm}
\usepackage{dcolumn}% Align table columns on decimal point
\begin{document}

\title{Excitation energies, polarizabilities, multipole transition rates,
and  lifetimes   in Th~IV }

\author{ U. I. Safronova }
\email{usafrono@nd.edu} \affiliation{Physics Department,
University of Nevada, Reno, NV
 89557}

\author{W. R. Johnson}
\email{johnson@nd.edu} \homepage{www.nd.edu/~johnson}
\affiliation{Department of Physics,  University of Notre Dame,
Notre Dame, IN 46556}

\author{M. S. Safronova}
\email{msafrono@udel.edu} \affiliation {Department of Physics and
Astronomy, 223 Sharp Lab,
 University of Delaware, Newark, Delaware 19716}

\date{\today}
\begin{abstract}

Excitation energies  of the  $ns_{1/2}$ ($n$= 7-10), $np_j$ ($n$ =
7-9), $nd_j$ ($n$= 6-8),  $nf_{j}$ ($n$= 5-7),  and $ng_{j}$ ($n$=
5-6) states in Th~IV  are evaluated. First-, second-, third-, and
all-order Coulomb energies and first- and second-order
Coulomb-Breit energies are calculated.  Reduced matrix elements,
oscillator strengths, transition rates, and lifetimes are
determined for the 96 possible $nl_j-n'l'_{j'}$ electric-dipole
transitions. Multipole  matrix elements ($7s_{1/2}\ - 6d_j$,
$7s_{1/2}\ - 5f_j$, and $5f_{5/2}\ - 5f_{7/2}$)  are evaluated to
obtain the lifetimes of the $5f_{7/2}$  and
$7s_{1/2}$ states.  Matrix elements are calculated using
both relativistic many-body perturbation theory, complete through
third order, and a relativistic all-order method restricted to
single and double (SD) excitations. Scalar and tensor
polarizabilities for the $5f_{5/2}$ ground state in Th$^{3+}$
 are calculated using
relativistic third-order and all-order methods. These calculations
provide a theoretical benchmark for comparison with experiment and
theory.
%31.15.Ar -Ab initio calculations
%31.15.Md - Perturbation theory
%32.10.Fn -Fine and hyperfine structure
%32.70.Cs -Oscillator strengths, lifetimes, transition moments
\end{abstract}
\maketitle

\section{Introduction}
A detailed investigation of radiative parameters for electric dipole
(E1) transitions in Fr-like ions with $Z$ = 89--92 was presented
recently by \citet{osc-ra}, where relativistic Hartree-Fock and
Dirac-Fock atomic structure codes were used to calculate
transition rates and oscillator strengths for a limited
number of transitions using experimental energies given by
Blaise and Wyart \cite{expt}.
In the compilation \cite{expt},
experimental energies are given for 56 levels of neutral Fr,
24 levels of Fr-like Th, and seven levels of Fr-like Ac
and U.  Experimental energies for 13 levels of Fr-like Ra were
reported in the NIST compilation \cite{web-nist}.

Lifetime measurements for the $7p_j$, $6d_j$,
$9s_{1/2}$, and $8s_{1/2}$ levels of neutral francium were presented in
Refs.~\cite{fr-97,fr-98,fr-00,fr-03,fr-05}. In those
papers, experimental measurements were compared with {\it
ab-initio} calculations performed by \citet{adndt-96},
by  \citet{dzuba-95,dzuba-01}, by \citet{safr-alk},
and by \citet{safr-fr}.
Third-order many-body perturbation theory was used in
Ref.~\cite{adndt-96} to obtain E1 transition amplitudes for neutral
alkali-metal atoms. The correlation potential method and the
Feynman diagram technique  was used in
Refs.~\cite{dzuba-95,dzuba-01} to calculate E1 dipole matrix
elements in neutral francium and in Fr-like radium. Calculations
of atomic properties of alkali-metal atoms in
Refs.~\cite{safr-alk,safr-fr} were based on the relativistic
single-double (SD) approximation in which single and double
excitations of Dirac-Fock wave functions were included in all
orders in perturbation theory.

In the present paper, relativistic many-body perturbation theory
(RMBPT) is used to determine energies, matrix elements, oscillator
strengths, and transition rates for multipole transitions in
Fr-like thorium. These calculations start from a radonlike closed-shell
Dirac-Fock (DF) potential.
It should be noted that Th~IV is the first ion in francium isoelectronic
sequence with a [Rn]$5f_{5/2}$ ground state instead of a [Rn]$7s_{1/2}$
ground state, as for Fr~I, Ra~II, and As~III. Correlation
corrections become very important for such systems as was
recently demonstrated in \citet{igor}, where it was shown that the ratio
of the second-order to lowest-order removal energy for the [Xe]$4f_{5/2}$
ground state in Ce~IV and Pr~V is 18~\% and 11~\%,
respectively.

We calculate excitation energies  of
$ns_{1/2}$ ($n$= 7-10), $np_j$ ($n$ = 7-9), $nd_j$ ($n$= 6-8),
$nf_{j}$ ($n$= 5-7),  and $ng_{j}$ ($n$= 5-6) states in Fr-like  thorium.
  Reduced matrix elements, oscillator strengths,
transition rates, and lifetimes are determined for the 96 possible
$nl_j-n'l'_{j'}$ electric-dipole transitions. Multipole  matrix
elements ($7s_{1/2}\ - 6d_j$, $7s_{1/2}\ - 5f_j$, and $5f_{5/2}\ -
5f_{7/2}$)  are evaluated to obtain the lifetimes of
$5f_{7/2}$ and  $7s_{1/2}$ states.
Scalar and tensor polarizabilities of the $5f_{5/2}$ ground
state of Th$^{3+}$  are also calculated.
Matrix elements are calculated using both
relativistic many-body perturbation theory, complete through third
order, and the relativistic all-order method restricted to single
and double (SD) excitations.
Such calculations permit one to investigate the convergence of perturbation
theory and estimate the error in theoretical data.

\begin{table*}
\caption{\label{tab1} Zeroth- (DF), second-, and third-order
Coulomb correlation energies $E^{(n)}$,  single-double Coulomb
energies $E^\text{{SD}}$, $E^{(3)}_\text{{extra}}$,
 first-order Breit and second-order Coulomb-Breit
 corrections $B^{(n)}$  to the energies of Th~IV.
The total energies ($E^{(3)}_\text{ tot} = E^{(0)} + E^{(2)}
+E^{(3)}+ B^{(1)} + B^{(2)} +E_{\rm LS}$, $E^\text{SD}_\text{tot}
= E^{(0)} + E^\text{{SD}} + E^{(3)}_\text{{extra}} + B^{(1)} +
B^{(2)} + E_{\rm LS}$) for Th~IV  are compared with experimental
 energies $E_\text{{expt}}$ \protect\cite{expt},
 $\delta E$ = $E_\text{tot}$ - $E_\text{{expt}}$. Units: cm$^{-1}$.}
\begin{ruledtabular}\begin{tabular}{lrrrrrrrrrrrrr}
\multicolumn{1}{c}{$nlj$ } & \multicolumn{1}{c}{$E^{(0)}$} &
\multicolumn{1}{c}{$E^{(2)}$} & \multicolumn{1}{c}{$E^{(3)}$} &
\multicolumn{1}{c}{$B^{(1)}$} & \multicolumn{1}{c}{$B^{(2)}$} &
\multicolumn{1}{c}{$E_{LS}$} & \multicolumn{1}{c}{$E^{(3)}_\text{
tot}$} & \multicolumn{1}{c}{$E^\text{{SD}}$} &
\multicolumn{1}{c}{$E^{(3)}_\text{{extra}}$} &
\multicolumn{1}{c}{$E^\text{{SD}}_\text{ tot}$} &
\multicolumn{1}{c}{$E_\text{{expt}}$} & \multicolumn{1}{c}{$\delta
E^{(3)}$} &
\multicolumn{1}{c}{$\delta E^\text{{SD}}$} \\
\hline
 $5f_{5/2}$& -206606& -32100&   11739&  704&  -2747&  0& -229010& -26327& 4672&  -230304& -231065&    2055&  761\\
 $5f_{7/2}$& -203182& -30549&   10954&  521&  -2616&  0& -224872& -25252& 4361&  -226168& -226740&    1868&  571\\
 $6d_{3/2}$& -211799& -13258&    4129&  438&   -880&  0& -221370& -11422& 1663&  -222000& -221872&     502& -128\\
 $6d_{5/2}$& -207574& -11608&    3300&  326&   -807&  0& -216364& -10208& 1337&  -216927& -216579&     214& -348\\
 $7s_{1/2}$& -200273& -11204&    4402&  325&   -458& 89& -207119&  -9455& 1697&  -208075& -207934&     815& -140\\
 $7p_{1/2}$& -165095&  -7991&    2782&  298&   -272&  0& -170278&  -7147& 1125&  -171091& -170826&     548& -265\\
 $7p_{3/2}$& -153572&  -6213&    2124&  184&   -226&  1& -157703&  -5619&  861&  -158372& -158009&     306& -363\\
 $8s_{1/2}$& -109201&  -3840&    1462&  122&   -163& 25& -111594&  -3255&  575&  -111896& -111443&    -151& -453\\
 $7d_{3/2}$& -108639&  -3815&    1110&  113&   -199&  0& -111429&  -3559&  496&  -111787& -111380&     -49& -406\\
 $7d_{5/2}$& -107032&  -3592&    1011&   86&   -188&  0& -109715&  -3401&  452&  -110083& -109638&     -77& -445\\
 $6f_{5/2}$&  -99921&  -5078&    1615&   77&   -270&  0& -103577&  -4763&  710&  -104167& -103796&     219& -371\\
 $6f_{7/2}$&  -99481&  -4998&    1590&   61&   -276&  0& -103103&  -4644&  696&  -103644& -103250&     146& -394\\
 $8p_{1/2}$&  -94597&  -3199&    1112&  129&   -116&  0&  -96670&  -2861&  460&   -96984&  -96549&    -122& -436\\
 $8p_{3/2}$&  -89595&  -2628&     898&   83&   -101&  0&  -91342&  -2397&  371&   -91638&  -91194&    -148& -444\\
 $5g_{7/2}$&  -70583&  -1506&     421&    0&     -4&  0&  -71672&  -1592&  207&   -71971&  -71694&      22& -296\\
 $5g_{9/2}$&  -70602&  -1491&     422&    0&     -4&  0&  -71675&  -1548&  205&   -71949&  -71675&       0& -255\\
 $9s_{1/2}$&  -69455&  -1849&     684&   61&    -80& 11&  -70628&  -1620&  275&   -70810&  -70337&    -291& -473\\
 $8d_{3/2}$&  -68677&  -1846&     522&   54&    -93&  0&  -70040&  -1821&  244&   -70292&  -69111&    -929&-1181\\
 $8d_{5/2}$&  -67864&  -1767&     488&   42&    -89&  0&  -69190&  -1780&  228&   -69463&  -68537&    -653& -926\\
 $7f_{5/2}$&  -64674&  -2576&     798&   40&   -137&  0&  -66549&  -2504&  362&   -66912&  -66510&     -39& -402\\
 $7f_{7/2}$&  -64454&  -2517&     782&   31&   -139&  0&  -66296&  -2409&  352&   -66619&  -66006&    -290& -613\\
 $9p_{1/2}$&  -61858&  -1663&     566&   68&    -61&  0&  -62948&  -1548&  241&   -63158&        &        &     \\
 $9p_{3/2}$&  -59200&  -1400&     468&   45&    -54&  0&  -60141&  -1352&  199&   -60362&        &        &     \\
 $6g_{7/2}$&  -49121&  -1042&     281&    0&     -4&  0&  -49884&  -1242&  145&   -50221&  -49391&    -493& -831\\
 $6g_{9/2}$&  -49138&  -1022&     285&    0&     -4&  0&  -49879&  -1166&  142&   -50166&  -49390&    -489& -775\\
$10s_{1/2}$&  -48177&  -1046&     376&   35&    -46&  5&  -48853&   -990&  154&   -49018&  -48624&    -229& -394\\
\end{tabular}
\end{ruledtabular}
\end{table*}

\begin{table*}
\caption{\label{tab-dip} Reduced electric-dipole matrix elements
 calculated to first, second, third, and all
orders of RMBPT
 in  Th~IV.}
\begin{ruledtabular}
\begin{tabular}{llrrrllrrr}
\multicolumn{2}{c}{Transition}& \multicolumn{1}{c}{$Z^{({\rm
DF})}$ }& \multicolumn{1}{c}{$Z^{({\rm DF}+2+3)}$ }&
\multicolumn{1}{c}{$Z^\text{{(SD)}}$ }&
\multicolumn{2}{c}{Transition}& \multicolumn{1}{c}{$Z^{({\rm
DF})}$ }& \multicolumn{1}{c}{$Z^{({\rm DF}+2+3)}$ }&
\multicolumn{1}{c}{$Z^\text{{(SD)}}$ }\\
\hline
 $ 5g_{7/2}$&$   5f_{5/2}$&  1.1236&   0.6400&   0.7118& $ 6d_{5/2}$&$   6f_{7/2}$&  3.3539&   3.1272&   3.0192\\
 $ 5g_{7/2}$&$   5f_{7/2}$&  0.2298&   0.1344&   0.1511& $ 6d_{5/2}$&$   7f_{5/2}$&  0.3586&   0.2561&   0.2398\\
 $ 5g_{7/2}$&$   6f_{5/2}$&  9.9381&   8.8278&   8.8815& $ 6d_{5/2}$&$   7f_{7/2}$&  1.5854&   1.1995&   1.1083\\
 $ 5g_{7/2}$&$   6f_{7/2}$&  1.9244&   1.7141&   1.7252& $ 6d_{5/2}$&$   7p_{3/2}$&  3.1975&   2.7006&   2.7549\\
 $ 5g_{7/2}$&$   7f_{5/2}$&  8.5105&   8.6350&   8.4527& $ 6d_{5/2}$&$   8p_{3/2}$&  0.6529&   0.4291&   0.4288\\
 $ 5g_{7/2}$&$   7f_{7/2}$&  1.6217&   1.6446&   1.6082& $ 6d_{5/2}$&$   9p_{3/2}$&  0.3496&   0.1743&   0.1438\\
            &             &        &         &         &            &             &        &         &         \\
 $ 6g_{7/2}$&$   5f_{5/2}$&  0.8677&   0.4157&   0.4134& $ 7d_{5/2}$&$   5f_{5/2}$&  0.0048&   0.0360&   0.0521\\
 $ 6g_{7/2}$&$   5f_{7/2}$&  0.1757&   0.0766&   0.0918& $ 7d_{5/2}$&$   5f_{7/2}$&  0.0703&   0.2298&   0.2961\\
 $ 6g_{7/2}$&$   6f_{5/2}$&  1.3517&   1.2696&   1.1321& $ 7d_{5/2}$&$   6f_{5/2}$&  2.0852&   1.8440&   1.8605\\
 $ 6g_{7/2}$&$   6f_{7/2}$&  0.2464&   0.2319&   0.2042& $ 7d_{5/2}$&$   6f_{7/2}$&  9.3944&   8.3249&   8.4004\\
 $ 6g_{7/2}$&$   7f_{5/2}$& 14.6139&  13.4006&  13.5178& $ 7d_{5/2}$&$   7f_{5/2}$&  1.0359&   1.1377&   1.1176\\
 $ 6g_{7/2}$&$   7f_{7/2}$&  2.8402&   2.6137&   2.6402& $ 7d_{5/2}$&$   7f_{7/2}$&  4.5090&   4.9585&   4.8615\\
            &             &        &         &         & $ 7d_{5/2}$&$   7p_{3/2}$&  5.9481&   5.3934&   5.4192\\
 $ 5g_{9/2}$&$   5f_{7/2}$&  1.3635&   0.8187&   0.9054& $ 7d_{5/2}$&$   8p_{3/2}$&  6.8642&   6.5224&   6.5180\\
 $ 5g_{9/2}$&$   6f_{7/2}$& 11.3858&  10.1422&  10.2068& $ 7d_{5/2}$&$   9p_{3/2}$&  0.8580&   0.7746&   0.7645\\
 $ 5g_{9/2}$&$   7f_{7/2}$&  9.5778&   9.7291&   9.5435&            &             &        &         &         \\
 $ 6g_{9/2}$&$   5f_{7/2}$&  1.0425&   0.5803&   0.5682& $ 8d_{5/2}$&$   5f_{5/2}$&  0.0062&   0.0307&   0.0461\\
 $ 6g_{9/2}$&$   6f_{7/2}$&  1.4474&   1.3738&   1.2252& $ 8d_{5/2}$&$   5f_{7/2}$&  0.0082&   0.2138&   0.2520\\
 $ 6g_{9/2}$&$   7f_{7/2}$& 16.8152&  15.4636&  15.5953& $ 8d_{5/2}$&$   6f_{5/2}$&  0.3783&   0.3187&   0.3357\\
            &             &        &         &         & $ 8d_{5/2}$&$   6f_{7/2}$&  1.7787&   1.5158&   1.6020\\
 $ 6d_{3/2}$&$   5f_{5/2}$&  2.4281&   1.3367&   1.5295& $ 8d_{5/2}$&$   7f_{5/2}$&  3.5961&   3.3715&   3.3634\\
 $ 6d_{3/2}$&$   6f_{5/2}$&  2.6761&   2.4407&   2.3610& $ 8d_{5/2}$&$   7f_{7/2}$& 16.1550&  15.1766&  15.1464\\
 $ 6d_{3/2}$&$   7f_{5/2}$&  1.2888&   0.9082&   0.8469& $ 8d_{5/2}$&$   7p_{3/2}$&  1.1741&   0.8883&   0.8651\\
 $ 6d_{3/2}$&$   7p_{1/2}$&  2.5465&   2.0723&   2.1220& $ 8d_{5/2}$&$   8p_{3/2}$&  9.3341&   8.9653&   8.9992\\
 $ 6d_{3/2}$&$   7p_{3/2}$&  0.9963&   0.8270&   0.8488& $ 8d_{5/2}$&$   9p_{3/2}$& 11.4942&  11.1245&  11.0183\\
 $ 6d_{3/2}$&$   8p_{1/2}$&  0.4074&   0.1809&   0.1906&            &             &        &         &         \\
 $ 6d_{3/2}$&$   8p_{3/2}$&  0.2173&   0.1410&   0.1422& $ 7s_{1/2}$&$   7p_{1/2}$&  2.8994&   2.3669&   2.4196\\
 $ 6d_{3/2}$&$   9p_{1/2}$&  0.2148&   0.0314&   0.0221& $ 7s_{1/2}$&$   7p_{3/2}$&  3.9933&   3.2930&   3.3677\\
 $ 6d_{3/2}$&$   9p_{3/2}$&  0.1172&   0.0568&   0.0475& $ 7s_{1/2}$&$   8p_{1/2}$&  0.0565&   0.2438&   0.2346\\
            &             &        &         &         & $ 7s_{1/2}$&$   8p_{3/2}$&  0.3273&   0.0639&   0.0679\\
 $ 7d_{3/2}$&$   5f_{5/2}$&  0.0654&   0.2134&   0.2587& $ 7s_{1/2}$&$   9p_{1/2}$&  0.0550&   0.1958&   0.1849\\
 $ 7d_{3/2}$&$   6f_{5/2}$&  7.8264&   6.9261&   6.9826& $ 7s_{1/2}$&$   9p_{3/2}$&  0.1326&   0.0706&   0.0674\\
 $ 7d_{3/2}$&$   7f_{5/2}$&  3.5114&   3.8940&   3.8365&            &             &        &         &         \\
 $ 7d_{3/2}$&$   7p_{1/2}$&  3.8261&   3.4234&   3.4490& $ 8s_{1/2}$&$   7p_{1/2}$&  1.5874&   1.5601&   1.5492\\
 $ 7d_{3/2}$&$   7p_{3/2}$&  2.0308&   1.8374&   1.8444& $ 8s_{1/2}$&$   7p_{3/2}$&  3.0768&   3.0010&   2.9756\\
 $ 7d_{3/2}$&$   8p_{1/2}$&  5.4788&   5.1788&   5.1791& $ 8s_{1/2}$&$   8p_{1/2}$&  5.0325&   4.6912&   4.7280\\
 $ 7d_{3/2}$&$   8p_{3/2}$&  2.1716&   2.0607&   2.0630& $ 8s_{1/2}$&$   8p_{3/2}$&  6.7737&   6.3299&   6.3880\\
 $ 7d_{3/2}$&$   9p_{1/2}$&  0.4173&   0.3208&   0.3214& $ 8s_{1/2}$&$   9p_{1/2}$&  0.0758&   0.1951&   0.2099\\
 $ 7d_{3/2}$&$   9p_{3/2}$&  0.3057&   0.2822&   0.2790& $ 8s_{1/2}$&$   9p_{3/2}$&  0.5353&   0.4009&   0.3719\\
            &             &        &         &         &            &             &        &         &         \\
 $ 8d_{3/2}$&$   5f_{5/2}$&  0.0029&   0.1874&   0.2098& $ 9s_{1/2}$&$   7p_{1/2}$&  0.4722&   0.4773&   0.4657\\
 $ 8d_{3/2}$&$   6f_{5/2}$&  1.6194&   1.3910&   1.4493& $ 9s_{1/2}$&$   7p_{3/2}$&  0.7567&   0.7254&   0.7123\\
 $ 8d_{3/2}$&$   7f_{5/2}$& 13.4659&  12.6494&  12.6151& $ 9s_{1/2}$&$   8p_{1/2}$&  2.8858&   2.8094&   2.8070\\
 $ 8d_{3/2}$&$   7p_{1/2}$&  0.9598&   0.7550&   0.7418& $ 9s_{1/2}$&$   8p_{3/2}$&  5.4176&   5.3126&   5.2854\\
 $ 8d_{3/2}$&$   7p_{3/2}$&  0.3604&   0.2581&   0.2523& $ 9s_{1/2}$&$   9p_{1/2}$&  7.6975&   7.4191&   7.4316\\
 $ 8d_{3/2}$&$   8p_{1/2}$&  5.9596&   5.6802&   5.7284& $ 9s_{1/2}$&$   9p_{3/2}$& 10.2453&   9.8671&   9.8961\\
 $ 8d_{3/2}$&$   8p_{3/2}$&  3.2096&   3.0818&   3.0900&            &             &        &         &         \\
 $ 8d_{3/2}$&$   9p_{1/2}$&  9.1563&   8.8665&   8.8044& $10s_{1/2}$&$   7p_{1/2}$&  0.2605&   0.2651&   0.2489\\
 $ 8d_{3/2}$&$   9p_{3/2}$&  3.6602&   3.5363&   3.5120& $10s_{1/2}$&$   7p_{3/2}$&  0.4011&   0.3784&   0.3596\\
            &             &        &         &         & $10s_{1/2}$&$   8p_{1/2}$&  0.7932&   0.7936&   0.7873\\
 $ 6d_{5/2}$&$   5f_{5/2}$&  0.6391&   0.3624&   0.4116& $10s_{1/2}$&$   8p_{3/2}$&  1.2032&   1.1808&   1.1676\\
 $ 6d_{5/2}$&$   5f_{7/2}$&  2.9557&   1.7032&   1.9190& $10s_{1/2}$&$   9p_{1/2}$&  4.5019&   4.3611&   4.3523\\
 $ 6d_{5/2}$&$   6f_{5/2}$&  0.7669&   0.7085&   0.6847& $10s_{1/2}$&$   9p_{3/2}$&  8.3091&   8.1560&   8.0743\\
\end{tabular}
\end{ruledtabular}
\end{table*}

\section{Third-order and all-order RMBPT calculations  of energies}

As mentioned in the introduction, we carry out all of the calculations in
this work using two methods,
third-order MBPT, described in \cite{adndt-96}, and the relativistic
all-order SD method, described in \cite{cs,relsd} and references therein.
The SD method includes correlation corrections in a more complete way and
is expected to yield more accurate results, especially when correlation
corrections are significant. While the SD method includes fourth-
and higher-order terms, it omits
some third-order terms. These omitted terms are identified and added to our SD
data (see \cite{relsd} for details).

We use the B-spline method \cite{Bspline} to generate a complete set of
basis DF orbitals for use in the evaluation of RMBPT
expressions.  For  Th~IV, we use 50 splines of order $k=8$ for each
angular momentum. The basis orbitals are constrained to a
spherical cavity of radius  $R$ = 45 a.u.. The cavity radius is
chosen large enough to accommodate all $nl_j$  orbitals considered
here and small enough that 50 splines can approximate inner-shell
DF wave functions with good precision. We use only 40 out of 50 basis
orbitals for each partial wave in our third-order and all-order
energy calculations since contributions from higher-energy
orbitals are negligible.

Results of our  energy calculations for the  28 states of Th~IV are summarized in
Table~\ref{tab1}. Columns 2--7 of Table~\ref{tab1} give the
lowest-order DF energies $E^{(0)}$, second- and third-order
Coulomb correlation energies, $E^{(2)}$ and $E^{(3)}$,
 first-order Breit contribution $B^{(1)}$,  second-order
Coulomb-Breit corrections  $B^{(2)}$, and  the Lamb shift $E_{\rm LS}$.
The sum of these six
contributions is our final third-order RMBPT result
$E^{(3)}_{\rm tot}$ listed in the eighth column of
Table~\ref{tab1}. The all-order SD energies are listed in the column
$E^\text{{SD}}$, and that part of the third-order energy
omitted in the SD calculation is given in column $E^{(3)}_\text{{extra}}$.
We note that $E^\text{{SD}}$ includes $E^{(2)}$ completely. We
take the sum of the six terms $E^{(0)}$, $E^\text{{SD}}$,
$E^{(3)}_\text{{extra}}$, $B^{(1)}$,  $B^{(2)}$, and  $E_{\rm LS}$
to be our final all-order result $E^\text{{SD}}_{\rm tot}$, listed
in the eleventh column of Table~\ref{tab1}.  Experimental
energies from Blaise and Wyart \cite{expt} are
given in the column labeled $E_\text{{expt}}$. Differences between
third-order and experimental energies $\delta E^{(3)}=E^{(3)}_{\rm tot}-E_\text{{expt}}$,
and between SD and experimental energies  $\delta E^\text{SD}=E^\text{{SD}}_{\rm tot}-E_\text{{expt}}$,
are given in the last two columns of Table~\ref{tab1}, respectively.

As expected, the largest correlation contribution to the valence
energy  comes from the second-order term $E^{(2)}$.
This term is  relatively simple to calculate; thus, we calculate
$E^{(2)}$ with better numerical accuracy than $E^{(3)}$ and $E^{\rm SD}$.
The second-order energy $E^{(2)}$ includes partial
waves up to $l_{\text{max}}=8$ and is extrapolated to account for
contributions from higher partial waves (see, for example,
\cite{be-en,be3-en}). As an example of the convergence of
$E^{(2)}$ with the number of partial waves $l$, consider the
$5f_{5/2}$ state in  Th~IV. Calculations of $E^{(2)}$ with
$l_{\text{max}}$ = 6 and  8 yield
 $E^{(2)}(5f_{5/2})$ = $-30810$ and
$-31771$~cm$^{-1}$, respectively. Extrapolation of these
calculations yields $-32100$ and $-32154$~cm$^{-1}$, respectively.
Therefore, we estimate the numerical uncertainty of
$E^{(2)}(5f_{5/2})$ to be approximately 54~cm$^{-1}$. It should be
noted that this is the largest contribution from the higher
partial waves, since we obtain a numerical uncertainty of   26~cm$^{-1}$
for $E^{(2)}(6d_{j})$ and the numerical
uncertainty of  1~cm$^{-1}$ for $E^{(2)}(7s_{j})$. Similar
convergence patterns are found for all other states considered.

We use $l_{\text{max}}$ = 6  in our all-order calculations owing
to the numerical complexity of the $E^{\rm SD}$ calculation. As we
noted above, the second-order $E^{(2)}$ is contained in the
$E^{\rm SD}$ value. Therefore, we use our high-precision
calculation of $E^{(2)}$ described above to account for the
contributions of the  higher partial waves. We simply replace
$E^{(2)}$[$l_{\text{max}}$ = 6] value with the final
high-precision second-order value  $E^{(2)}_{\rm final}$. The same
number of partial waves, $l_{\text{max}}$ = 6, is used in the
third-order calculation.
  Since the asymptotic
$l$-dependence of the second- and third-order energies are similar
(both fall off as $l^{-4}$), we use the second-order remainder as
a guide to estimate the numerical errors in  the third-order
contribution. The contribution $E^{(3)}_\text{{extra}}$ given
 in Table~\ref{tab1} accounts for that part of the
third-order RMBPT correction  not included in the SD
energy. The values of $E^{(3)}_\text{{extra}}$ are smaller than
the values of $E^{(3)}$ by approximately  a factor of 3.

 The first-order Breit energies (column $B^{(1)}$ of Table~\ref{tab1})  include
retardation, whereas the second-order Coulomb-Breit energies
(column $B^{(2)}$ of Table~\ref{tab1}) are evaluated using the
unretarded Breit operator. The total $E^{(3)}_\text{ tot}$ in
Table~\ref{tab1} is the sum of six terms, $E^{(0)}$, $E^{(2)}$,
$E^{(3)}$, $B^{(1)}$, $B^{(2)}$, and $E_{\rm LS}$. We find that
the correlation corrections to energies  are large, especially for
the $5f_j$ states.  For example, $E^{(2)}$ is about 15\% of
$E^{(0)}$ and $E^{(3)}$ is about 36\% of $E^{(2)}$ for the
$5f_{j}$ states. Despite the evident slow convergence of the
perturbation theory expansion, the $5f_{j}$ energy from the
third-order RMBPT calculation is within 0.9\% of the measured
energy. It should be noted that correlation corrections are much
smaller for all other states; the ratios of $E^{(0)}$ and
$E^{(2)}$ are equal to 6\%, 5\%, and 2\%  for the $6d_{j}$,
$7s_{1/2}$, and $10s_{1/2}$ states, respectively.
An important consequence of the large size of correlation corrections for $5f_{j}$ states
is a different ordering of uncorrelated and correlated
energies.  As can be seen from Table~\ref{tab1}, $-E^{(0)}$ values
for $6d_{j}$ states are larger than $-E^{(0)}$ values
for $5f_{j}$ states; however,  $-(E^{(0)} + E^{(2)})$
values for $6d_{j}$ states are smaller than  $-(E^{(0)} + E^{(2)})$
values for $5f_{j}$ states; thus, although DF calculations predict the
ground state of Th~IV to be $6d_{3/2}$, correlated calculations correctly
predict the ground state to be $5f_{5/2}$.

The quantity $E^\text{SD}_\text{ tot}$ in Table~\ref{tab1} is the sum
of six terms; $E^{(0)}$, $E^\text{{SD}}$,
$E^{(3)}_\text{{extra}}$, $B^{(1)}$,  $B^{(2)}$, and $E^{LS)}$.
The column labeled $\delta E^\text{{SD}}$ in Table~\ref{tab1}
gives differences between our {\it ab initio}
 results and the experimental values \cite{expt}.
The SD results agree better with the experimental values than  the
third-order RMBPT results for low-lying states where the correlation
correction is larger. Comparison of the results given in two
last columns of Table~\ref{tab1} shows that the ratio of $\delta
E^{(3)}$ and $\delta E^\text{{SD}}$ is about 3 for the $5f_{j}$
states. As expected, including correlation to all orders led to
significant improvement of the results. Better agreement of the
all-order values with experiment demonstrates the importance of
the higher order correlation contributions.

\section{Electric-dipole matrix elements, oscillator strengths, transition
rates, and lifetimes in Th~IV}

\subsection{Electric-dipole matrix elements}

The calculation of the transition matrix elements provides another
test of the quality of atomic-structure calculations and another
measure of the size of correlation corrections. Reduced
electric-dipole matrix elements between low-lying states of Th~IV
 calculated in third order RMBPT and in the
 SD approximation are presented in Table~\ref{tab-dip}.

Third-order matrix elements $Z^{({\rm DF}+2+3)}$ include DF contributions together with
second-order $Z^{(2)}$ and third-order $Z^{(3)}$  correlation corrections.
Second- and third-order random-phase-approximation (RPA) terms are iterated to all
orders in the present calculation. Third-order corrections include Brueckner orbital (BO),
structural radiation $Z^\text{(SR)}$,  and normalization $Z^{(\rm NORM)}$ corrections, in addition to
the third-order RPA terms, see \cite{adndt-96}.
The terms $Z^\text{ (RPA)}$ and $Z^\text{ (BO)}$ give the largest
contributions to the total.  The sum of terms $Z^\text{ (RPA)}$
and $Z^\text{ (BO)}$ is about 15--25\% of $Z^{({\rm DF})}$
and has a different sign.  Structural radiation,
 and normalization  corrections are
small. We find correlation corrections  $Z^{(2+3)}$ to be  very
large, 10-25\%, for many cases. All results given in
Table~\ref{tab-dip} are obtained using length-form  matrix
elements. Length-form and velocity-form matrix elements differ
typically by 5--20~\% for DF matrix elements and 2--5 \% for
the second-order matrix elements in these calculations.

Electric-dipole matrix elements evaluated in the all-order SD
approximation are given in  columns labeled $Z^\text{(SD)}$
 of Table~\ref{tab-dip}. The SD matrix elements
$Z^\text{(SD)}$ include $Z^{(3)}$ completely, along with important
fourth- and higher-order corrections.  The fourth-order corrections
omitted from the SD matrix elements were discussed recently by
\citet{der-4}.  The SD matrix elements $Z^\text{{(SD)}}$ are smaller than
$Z^{({\rm DF}+2)}$  but larger than $Z^{({\rm DF}+2+3)}$
for all of the transitions listed in Table~\ref{tab-dip}.

\begin{table}
\caption{\label{tab-LV} Comparison of length [L] and velocity [V]
results for reduced electric-dipole matrix elements in lowest and
third orders of perturbation theory in  Th~IV. }
\begin{ruledtabular}
\begin{tabular}{llrrrr}
\multicolumn{2}{c}{Transition}& \multicolumn{2}{c}{$Z^{({\rm
DF})}$ }&
\multicolumn{2}{c}{$Z^{({\rm DF}+2+3)}$}\\
\multicolumn{2}{c}{ }& \multicolumn{1}{c}{ $L$}&
\multicolumn{1}{c}{ $V$}& \multicolumn{1}{c}{ $L$}&
\multicolumn{1}{c}{ $V$}\\
\hline
 $ 5g_{7/2}$&$   5f_{5/2}$&    -1.1236&     -1.0851&       -0.7094&     -0.7080\\
 $ 5g_{7/2}$&$   5f_{7/2}$&     0.2298&      0.2211&        0.1479&      0.1476\\
 $ 5g_{7/2}$&$   6f_{5/2}$&     9.9381&      9.9552&        8.9038&      8.8996\\
 $ 6d_{3/2}$&$   5f_{5/2}$&    -2.4281&     -0.7698&       -2.8562&     -2.9000\\
 $ 7d_{3/2}$&$   5f_{5/2}$&    -0.0654&     -0.0403&       -0.1829&     -0.1810\\
 $ 7d_{3/2}$&$   9p_{3/2}$&     0.3057&      0.2702&        0.2831&      0.2830\\
 $ 8d_{3/2}$&$   5f_{5/2}$&     0.0029&      0.0123&       -0.1588&     -0.1579\\
 $ 6d_{5/2}$&$   5f_{5/2}$&     0.6391&     -2.2929&        0.6529&      0.6615\\
 $ 6d_{5/2}$&$   5f_{7/2}$&     2.9557&      0.5609&        3.3454&      3.3138\\
 $ 7d_{5/2}$&$   5f_{5/2}$&    -0.0048&      0.0017&       -0.0309&     -0.0303\\
 $ 7d_{5/2}$&$   5f_{7/2}$&    -0.0703&     -0.0425&       -0.2007&     -0.2025\\
 $ 7d_{5/2}$&$   6f_{5/2}$&     2.0852&      2.3007&        1.7790&      1.7779\\
 $ 8d_{5/2}$&$   5f_{7/2}$&    -0.0082&     -0.0178&        0.1823&      0.1833\\
 $ 8d_{5/2}$&$   6f_{5/2}$&     0.3783&      0.4084&        0.3258&      0.3256\\
 $ 7s_{1/2}$&$   7p_{1/2}$&     2.8994&      2.6904&        2.3943&      2.3936\\
 $ 7s_{1/2}$&$   7p_{3/2}$&    -3.9933&     -3.7030&       -3.3434&     -3.3432\\
 $ 7s_{1/2}$&$   8p_{1/2}$&    -0.0565&     -0.0910&       -0.2334&     -0.2335\\
 $ 8s_{1/2}$&$   9p_{1/2}$&    -0.0758&     -0.1060&       -0.1931&     -0.1932\\
 $ 8s_{1/2}$&$   9p_{3/2}$&     0.5353&      0.4761&        0.4031&      0.4030\\
 $ 9s_{1/2}$&$   7p_{1/2}$&    -0.4722&     -0.4420&       -0.4737&     -0.4736\\
 $ 9s_{1/2}$&$   7p_{3/2}$&     0.7567&      0.6966&        0.7181&      0.7180\\
 $ 9s_{1/2}$&$   8p_{1/2}$&    -2.8858&     -2.8249&       -2.8139&     -2.8137\\
 \end{tabular}
\end{ruledtabular}
\end{table}

\subsection{Form-independent third-order transition amplitudes}
We calculate electric-dipole reduced matrix elements using the
form-independent third-order perturbation theory developed by
Savukov and Johnson in Ref.~\cite{savukov01}. The
precision of this method has been demonstrated previously for alkali-metal
atoms.  In this method, form-dependent ``bare''  amplitudes are
 replaced with  form-independent
random-phase approximation (``dressed'') amplitudes to obtain
form-independent third-order amplitudes.
As in the case of the third-order energy calculation, a
limited number of partial waves with $l_{\max }<7$ is included,
 giving rise to some loss of gauge invariance.
Comparison of length- and velocity-form matrix elements serves as
a measure of the numerical accuracy of the resulting calculations.

Length- and velocity-form matrix elements from DF, second-order,
 and third-order calculations are given in
Table~\ref{tab-LV} for the limited number transitions in
Th~IV.
 Following the procedure discussed in Ref.~\cite{savukov01}, the DF
and RPA matrix elements in the table were obtained by dividing the
corresponding amplitude by the lowest-order transition energies
while the third-order matrix elements were obtained by
dividing the third-order amplitude by the second-order transition
energies. Values of $Z^{(\rm DF)}$ differ in $L$ and $V$ forms by
2--15\% for the $p$-$s$ transitions. Huge $L-V$ differences in the
 $Z^{(\rm DF)}$ for $d$-$f$ transitions can be seen in Table~\ref{tab-LV}.
Third-order calculations essentially remove such differences;
the residual differences (0.002\%--0.2\%) being explained
by the limited number of partial waves used in the evaluation
of third-order matrix elements.

\begin{table}
\caption{\label{tab-com1} Wavelengths $\lambda$ (\AA), weighted
transition rates $gA$ (s$^{-1}$), and oscillator strengths $gf$ in
Th~IV. The SD data
 ($gA^{\mathrm{(SD)}}$ and $gf^{\mathrm{(SD)}}$) are compared with
  theoretical ($gA^{\mathrm{(HFR)}}$ and $gf^{\mathrm{(HFR)}}$) values
  given in Ref.~\protect\cite{osc-ra}.  Numbers in brackets represent powers of 10. }
\begin{ruledtabular}
\begin{tabular}{llrllll}
\multicolumn{1}{c}{Lower}& \multicolumn{1}{c}{Upper}&
\multicolumn{1}{c}{$\lambda^\text{{(expt)}}$}&
\multicolumn{1}{c}{$gA^{\mathrm{(SD)}}$ }&
\multicolumn{1}{c}{$gA^{\mathrm{(HFR)}}$ }&
\multicolumn{1}{c}{$gf^{\mathrm{(SD)}}$ }&
\multicolumn{1}{c}{$gf^{\mathrm{(HFR)}}$ }\\
\hline
 $7s_{1/2}$ &   $7p_{3/2}$ &    2003.00&   2.86[+9]&    2.70[+9]&    1.72[+0]&    1.6[+0]\\
 $7s_{1/2}$ &   $7p_{1/2}$ &    2694.81&   6.06[+8]&    5.55[+8]&    6.60[-1]&    6.0[-1]\\[0.3pc]
 $7p_{1/2}$ &   $8s_{1/2}$ &    1684.00&   1.02[+9]&    1.62[+9]&    4.33[-1]&    6.9[-1]\\
 $7p_{3/2}$ &   $8s_{1/2}$ &    2147.50&   1.81[+9]&    1.56[+9]&    1.25[+0]&    1.1[+0]\\[0.3pc]
 $8s_{1/2}$ &   $8p_{3/2}$ &    4938.44&   6.87[+8]&    7.06[+8]&    2.51[+0]&    2.6[+0]\\
 $8s_{1/2}$ &   $8p_{1/2}$ &    6713.71&   1.50[+8]&    1.40[+8]&    1.01[+0]&    9.5[-1]\\[0.3pc]
 $6d_{3/2}$ &   $7p_{3/2}$ &    1565.86&   3.80[+8]&    4.08[+8]&    1.40[-1]&    1.5[-1]\\
 $6d_{5/2}$ &   $7p_{3/2}$ &    1707.37&   3.09[+9]&    2.83[+9]&    1.35[+0]&    1.2[+0]\\
 $6d_{3/2}$ &   $7p_{1/2}$ &    1959.02&   1.21[+9]&    1.04[+9]&    6.98[-1]&    6.0[-1]\\[0.3pc]
 $7p_{1/2}$ &   $7d_{3/2}$ &    1682.21&   5.06[+9]&    6.36[+9]&    2.15[+0]&    2.7[+0]\\
 $7p_{3/2}$ &   $7d_{5/2}$ &    2067.35&   6.73[+9]&    6.16[+9]&    4.32[+0]&    4.0[+0]\\
 $7p_{3/2}$ &   $7d_{3/2}$ &    2144.60&   6.99[+8]&    6.13[+8]&    4.82[-1]&    4.3[-1]\\[0.3pc]
 $7d_{3/2}$ &   $8p_{3/2}$ &    4953.85&   7.09[+7]&    8.03[+7]&    2.61[-1]&    3.0[-1]\\
 $7d_{5/2}$ &   $8p_{3/2}$ &    5421.88&   5.40[+8]&    5.51[+8]&    2.38[+0]&    2.5[+0]\\
 $7d_{3/2}$ &   $8p_{1/2}$ &    6742.22&   1.77[+8]&    1.59[+8]&    1.21[+0]&    1.1[+0]\\[0.3pc]
 $5f_{5/2}$ &   $6g_{7/2}$ &    550.433&   2.08[+9]&    7.40[+9]&    9.43[-2]&    3.4[-1]\\
 $5f_{7/2}$ &   $6g_{7/2}$ &    563.858&   9.51[+7]&    2.55[+8]&    4.54[-3]&    1.2[-2]\\
 $5f_{7/2}$ &   $6g_{9/2}$ &    563.861&   3.65[+9]&    8.93[+9]&    1.74[-1]&    4.3[-1]\\[0.3pc]
 $5f_{5/2}$ &   $5g_{7/2}$ &    627.392&   4.16[+9]&    8.14[+9]&    2.45[-1]&    4.8[-1]\\
 $5f_{7/2}$ &   $5g_{7/2}$ &    644.892&   1.72[+8]&    2.78[+8]&    1.08[-2]&    1.7[-2]\\
 $5f_{7/2}$ &   $5g_{9/2}$ &    644.971&   6.19[+9]&    9.72[+9]&    3.86[-1]&    6.0[-1]\\[0.3pc]
 $5f_{5/2}$ &   $6d_{5/2}$ &    6903.05&   1.04[+6]&    2.22[+6]&    7.45[-3]&    1.6[-2]\\
 $5f_{7/2}$ &   $6d_{5/2}$ &    9841.54&   7.83[+6]&    1.53[+7]&    1.14[-1]&    2.2[-1]\\
 $5f_{5/2}$ &   $6d_{3/2}$ &    10877.6&   3.68[+6]&    7.93[+6]&    6.53[-2]&    1.4[-1]\\
\end{tabular}
\end{ruledtabular}
\end{table}

\begin{table}
\caption{\label{tab-life} Lifetimes ${\tau}$ in ns
 for the $nl$  levels
 in Fr-like Th.}
\begin{ruledtabular}
\begin{tabular}{llll}
\multicolumn{1}{c}{Level}              &
\multicolumn{1}{c}{$\tau^{(\rm SD)}$}  & \multicolumn{1}{c}{Level}
& \multicolumn{1}{c}{$\tau^{(\rm SD)}$} \\\hline
   $6d_{3/2}$&     1090 &   $7p_{1/2}$&     1.099 \\
   $6d_{5/2}$&     676. &   $7p_{3/2}$&     0.632\\
   $7d_{3/2}$&     0.667&   $8p_{1/2}$&     3.194 \\
   $7d_{5/2}$&     0.854&   $8p_{3/2}$&     1.871 \\
   $8d_{3/2}$&     1.176&   $9p_{1/2}$&     5.893 \\
   $8d_{5/2}$&     1.600&   $9p_{3/2}$&     4.933 \\[0.4pc]
   $5g_{7/2}$&     0.815&   $6f_{5/2}$&     0.300\\
   $5g_{9/2}$&     0.780&   $6f_{7/2}$&     0.297\\
   $6g_{7/2}$&     1.768&   $7f_{5/2}$&     0.684\\
   $6g_{9/2}$&     1.567&   $7f_{7/2}$&     0.639\\[0.4pc]
   $8s_{1/2}$&     0.707& &\\
   $9s_{1/2}$&     1.031& &\\
   $10s_{1/2}$&    1.634& &\\
\end{tabular}
\end{ruledtabular}
\end{table}

\subsection{Oscillator strengths, transition rates and lifetimes}

We calculate  oscillator strengths and transition probabilities
for 96 possible $nl_j-n'l'_{j'}$ electric-dipole transitions
including the  $ns_{1/2}$ ($n$= 7-10), $np_j$ ($n$ = 7-9),
$nd_j$ ($n$= 6-8),  $nf_{j}$ ($n$= 5-7),  and $ng_{j}$ ($n$= 5-6)
states in Fr-like  thorium. Our results are presented in
Table~\ref{tab-com1} and Table~\ref{tab-life}.
Wavelengths $\lambda$ (\AA), weighted transition rates $gA$
(s$^{-1}$), and oscillator strengths $gf$ in Th~IV are given in
Table~\ref{tab-com1}. Our SD data,
 $gA^{\mathrm{(SD)}}$ and $gf^{\mathrm{(SD)}}$, are compared with
  theoretical calculations, $gA^{\mathrm{(HFR)}}$ and $gf^{\mathrm{(HFR)}}$,
  from Ref.~\cite{osc-ra}. It should be noted that
  experimental energies are used
  to calculate  $gA^{\mathrm{(SD)}}$ and $gf^{\mathrm{(SD)}}$ as well as
$gA^{\mathrm{(HFR)}}$ and $gf^{\mathrm{(HFR)}}$. Therefore,
we really compare the dipole matrix elements (see
Table~\ref{tab-dip}).
The  SD and HFR results for $s-p$
and  $p-d$ transitions disagree by  6--25\%, except for the
$7p_{1/2} - 8s_{1/2}$ transition  with  60\% disagreement. There
are also substantial disagreements  (factors of 2--5) between SD and
 HFR results for the $f-g$ and $f-d$ transitions.
 Correlation corrections are very important for those transitions as
discussed above (see Table~\ref{tab-dip}). The RPA and
BO contributions have the same sign opposite to the DF contributions
 and the
total values are half of the DF values. We see from
Table~\ref{tab-com1} that for the $f-g$ and $f-d$ transitions  the
values of $gA^{\mathrm{(HFR)}}$ and $gf^{\mathrm{(HFR)}}$ are
larger by a factor of 2--5 than the
 $gA^{\mathrm{(SD)}}$ and $gf^{\mathrm{(SD)}}$ values,
 respectively. On the basis of these comparisons, it appears that correlation
 corrections were not included in Ref.~\cite{osc-ra} for  transitions involving
  the $5f_j$ states. Our conclusion is confirmed by
 comparison of  $gA^{\mathrm{(HFR)}}$ and $gf^{\mathrm{(HFR)}}$
 with our $gA^{\mathrm{(DF)}}$ and $gf^{\mathrm{(DF)}}$ results
 (compare the $Z^{\rm DF}$ and $Z^{\rm SD}$ columns
 in Table ~\ref{tab-dip}).
 The disagreement between HFR and DF values for transitions rates
 and oscillator strengths is significantly smaller (about 10--20~\%) than the
 disagreement between HFR and SD  (by a factor of 2--5).

We calculate lifetimes of  $ns_{1/2}$ ($n$= 8-10), $np_j$ ($n$ =
7-9), $nd_j$ ($n$= 6-8),  $nf_{j}$ ($n$= 6-7),  and $ng_{j}$ ($n$=
5-6) states in Fr-like  thorium using the SD results for dipole
matrix elements and experimental energies~\cite{expt}. We list
lifetimes $\tau^\text{(SD)}$  in Table~\ref{tab-life}.
Unfortunately, there are no experimental measurements to compare
with our results; however, we hope that our calculations provide a
theoretical benchmark and lifetime measurements will be carried out.

Lifetimes for two excited levels,
$7s_{1/2}$ and $5f_{7/2}$ were not included in
Table~\ref{tab-life} since there are no electric-dipole
transitions from these levels. Contributions of the electric- and
magnetic-multipole transitions to the lifetime of the $7s_{1/2}$ and
$5f_{7/2}$ levels are considered below.

\begin{table*}
\caption{\label{tab-mult} Reduced matrix elements of the electric
multipole (E2, E3)
 and magnetic-multipole (M1, M2, and M3) operators in first, second,
third, and all orders of perturbation theory in  Th~IV.}
\begin{ruledtabular}
\begin{tabular}{lllrrrr}
\multicolumn{1}{c}{}& \multicolumn{2}{c}{Transition}&
\multicolumn{1}{c}{$Z^{({\rm DF})}$ }&
\multicolumn{1}{c}{$Z^{({\rm DF}+2)}$ }&
\multicolumn{1}{c}{$Z^{({\rm DF}+2+3)}$ }&
\multicolumn{1}{c}{$Z^\text{{(SD)}}$ }\\
\hline
M1   &$5f_{5/2}$ &$   5f_{7/2}$ &      1.8506&      1.8525&      1.8390&      1.8514\\
E2   &$5f_{5/2}$ &$   5f_{7/2}$ &      1.5669&      1.2339&      0.9724&      1.0834\\[0.4pc]
E2   &$6d_{3/2}$ &$   7s_{1/2}$ &      7.7806&      7.5518&      6.9232&      7.0631\\
E2   &$6d_{5/2}$ &$   7s_{1/2}$ &     10.0084&      9.7894&      8.9992&      9.1526\\
E3   &$5f_{5/2}$ &$   7s_{1/2}$ &     12.9552&     13.1718&      9.3847&     10.6349\\
E3   &$5f_{7/2}$ &$   7s_{1/2}$ &     15.7681&     16.0030&     11.5616&     13.0648\\
M3   &$6d_{5/2}$ &$   7s_{1/2}$ &     59.2100&     60.9891&     59.7136&     57.3087\\
\end{tabular}
\end{ruledtabular}
\end{table*}

\begin{table}
\caption{\label{tab-ar-mult} Wavelengths $\lambda$ (\AA) and
transition rates $A^{(SD)}$
 (s$^{-1}$) of the electric multipole (E2, E3)
 and magnetic-multipole M1, M2, and M3)
  transitions in Th~IV  calculated in
the SD approximation. Numbers in brackets represent powers of 10.}
\begin{ruledtabular}
\begin{tabular}{lllrl}
\multicolumn{1}{c}{} & \multicolumn{2}{c}{Transition} &
\multicolumn{1}{c}{$\lambda$} &
\multicolumn{1}{c}{$A^{\rm (SD)}$} \\
\hline
M1   &$5f_{5/2}$ &$   5f_{7/2}$ &     23119.6&     9.352[-1]\\
E2   &$5f_{5/2}$ &$   5f_{7/2}$ &     23119.6&     2.487[-5]\\[0.4pc]
E2   &$6d_{3/2}$ &$   7s_{1/2}$ &     7174.88&     1.469[-0]\\
E2   &$6d_{5/2}$ &$   7s_{1/2}$ &     11568.2&     2.264[-1]\\
M1   &$6d_{3/2}$ &$   7s_{1/2}$ &     7174.88&     5.429[-5]\\
E3   &$5f_{5/2}$ &$   7s_{1/2}$ &     4323.25&     6.299[-7]\\
E3   &$5f_{7/2}$ &$   7s_{1/2}$ &     5317.62&     2.232[-7]\\
M2   &$5f_{5/2}$ &$   7s_{1/2}$ &     4323.25&     5.154[-9]\\
M3   &$6d_{5/2}$ &$   7s_{1/2}$ &     11568.2&     1.863[-8]\\
\end{tabular}
\end{ruledtabular}
\end{table}

\section{Multipole matrix elements, transition rates, and
lifetimes in Th~IV}

Reduced matrix elements of the electric-quadrupole (E2),
electric-octupole (E3), and magnetic-multipole (M1, M2, and M3)
operators in lowest, second, third, and all orders of perturbation
theory are given in Table~\ref{tab-mult} for Th~IV. Detailed
description of the calculations  of the  multipole matrix elements
 in lowest and second orders of
perturbation theory were given  in
Refs.~\cite{multipol,ca,cjp-04}. Third-order and all-order calculations are
done in the same way as the calculations of the E1 matrix elements. In
Table~\ref{tab-mult}, we present  E2, E3, M1, M2, and M3 matrix
elements in the $Z^{(\rm DF)}$, $Z^{(\rm DF+2)}$, $Z^{(\rm
DF+2+3)}$, and $Z^\text{(SD)}$ approximations for the
$5f_{5/2}-5f_{7/2}$, $5f_{j}-7s_{1/2}$,
 and $6d_{j}-7s_{1/2}$ transitions in Th~IV.

The second-order
contribution is about 1--3\% for all transitions involving the
$7s_{1/2}$ states, but it is different for the $5f_{5/2} -
5f_{7/2}$ transition. It is very small (0.1\%) for the M1
matrix elements and rather
large (20\%) for the E2 matrix elements. The large difference between
$Z^{(\rm DF)}$ and $Z^{(\rm DF+2+3)}$ or  $Z^\text{(SD)}$ for E2
and E3 operators could be explained by the large size of
the Brueckner orbital (BO) correction; the ratios of $Z^{\rm (BO)}$
and $Z^{\rm (DF)}$ are equal to 0.06 and 0.3 for the
$6d_{j}-7s_{1/2}$ and $5f_{j}-7s_{1/2}$ transitions, respectively.

 Wavelengths and transition rates $A^{\textrm{(SD)}}$
 for the electric multipole (E2 and E3) and magnetic-multipole (M1, M2, and M3)
  transitions in Th~IV  calculated in the SD approximation are
  presented in Table~\ref{tab-ar-mult}. The largest contribution to the lifetime of the $5f_{7/2}$
  state comes from the M1 transition. The largest contribution
  to the lifetime of the $7s_{1/2}$
  state  comes from the E2 transitions. Our SD result for M1
  matrix elements are in perfect agreement (0.5\%) with HFR results obtained by
by Bi\'{e}mont {\it et al.\/} in Ref.~\cite{osc-ra}.
The disagreement is much larger between HFR and SD results for E2 transitions; 6\%
and 18\% for the $6d_{5/2} - 7s_{1/2}$ and $6d_{5/2} -
7s_{1/2}$ matrix elements, respectively. This is expected because
the correlation correction to the M1 matrix element is very small, but the
correlation correction to E2 matrix element is large, as discussed above.
Since the SD all-order method includes the correlation corrections in a
rather complete way, we expect to see disagreements with HFR
calculations in the cases where correlation corrections are significant.

Finally, we find that the lifetime of the  $5f_{7/2}$ state is
 1.07~s and the lifetime of the  $7s_{1/2}$ state is
 0.590~s. An estimate of the  $7s_{1/2}$ state lifetime (about
1~s) was given by  Peik and  Tamm in Ref.~\cite{life-03}, but
no measurements have as yet been performed.

\begin{table*}
\caption{\label{tab-comp2} Contributions to tensor polarizability
of Th~IV in the ground state $5f_{5/2}$ calculated using
third-order MBPT $\alpha_2^{\rm (DF+2+3)}$ and all-order SD method
$\alpha_2^{\rm (SD)}$. The third-order  and SD dipole matrix
elements and corresponding experimental transition energies are
also given. All values are in a.u.}
\begin{ruledtabular}
\begin{tabular}{lrrrrrrr}
\multicolumn{1}{c}{Contribution}& \multicolumn{1}{c}{$v$}&
\multicolumn{1}{c}{$n$}& \multicolumn{1}{c}{$E_{n}-E_{v} $ }&
\multicolumn{1}{c}{$Z_{vn}^{(\rm DF+2+3)}$ }&
\multicolumn{1}{c}{$Z_{vn}^{(\rm SD)}$ }&
\multicolumn{1}{c}{$\alpha_2^{\rm (DF+2+3)}$ }&
\multicolumn{1}{c}{$\alpha_2^{\rm (SD)}$ }\\
\hline
$\alpha_2^{\rm main}$  &$5f_{5/2} $&$  6d_{3/2} $&    0.041888&    1.337&  -1.530&   -4.740&    -6.206\\
&$5f_{5/2} $&$  7d_{3/2} $&    0.545323&    0.213&  -0.259&   -0.009&    -0.014   \\
&$5f_{5/2} $&$  8d_{3/2} $&    0.737917&    0.187&  -0.210&
-0.005&    -0.007   \\ [0.5pc]
&$5f_{5/2} $&$  6d_{5/2} $&    0.066005&    0.362&   0.412&    0.253&     0.326   \\
&$5f_{5/2} $&$  7d_{5/2} $&    0.553263&   -0.036&  -0.052&    0.000&     0.001\\
&$5f_{5/2} $&$  8d_{5/2} $&    0.740531&    0.031&   0.046&    0.000&     0.000\\[0.5pc]
&$5f_{5/2} $&$  5g_{7/2} $&    0.726234&    0.640&  -0.712&   -0.022&    -0.028\\
&$5f_{5/2} $&$  6g_{7/2} $&    0.827772&   -0.416&   0.413&   -0.008&    -0.008\\[0.5pc]
$\delta \alpha_2^{\rm core}\textrm{(DF)}$&&&&&                          &0.042& 0.042\\
$\alpha_2^{\rm tail}\textrm{(DF)}$  &&&&&&-0.273&-0.273\\
Total  &&&&& &-4.763 &-6.166
  \end{tabular}
\end{ruledtabular}
\end{table*}

\begin{table*}
\caption{\label{tab-comp6} Contributions to scalar polarizability
of Th~IV in the ground state $5f_{5/2}$ calculated using
third-order MBPT $\alpha_0^{\rm (DF+2+3)}$ and all-order SD method
$\alpha_0^{\rm (SD)}$. The third-order  and SD dipole matrix
elements and corresponding experimental transition energies are
also given. All values are in a.u. }
\begin{ruledtabular}
\begin{tabular}{lrrrrrrr}
\multicolumn{1}{c}{Contribution}& \multicolumn{1}{c}{$v$}&
\multicolumn{1}{c}{$n$}& \multicolumn{1}{c}{$E_{n}-E_{v} $ }&
\multicolumn{1}{c}{$Z_{vn}^{(\rm DF+2+3)}$ }&
\multicolumn{1}{c}{$Z_{vn}^{(\rm SD)}$ }&
\multicolumn{1}{c}{$\alpha_0^{\rm (DF+2+3)}$ }&
\multicolumn{1}{c}{$\alpha_0^{\rm (SD)}$ }\\
\hline
$\alpha_0^{\rm main}$ &$5f_{5/2} $&$  6d_{3/2} $&  0.041888&  1.337&  -1.530&    4.740&    6.206\\
&$5f_{5/2} $&$  6d_{5/2} $&  0.066005&  0.362&   0.412&    0.221&    0.285\\[0.5pc]
&$5f_{5/2} $&$  5g_{7/2} $&  0.726234&  0.640&  -0.712&    0.063&    0.078\\
&$5f_{5/2} $&$  6g_{7/2} $&  0.827772& -0.416&   0.413&    0.023&    0.023\\[0.5pc]
&$5f_{5/2} $&$  7d_{3/2} $&  0.545323&  0.213&  -0.259&    0.009&    0.014\\
&$5f_{5/2} $&$  8d_{3/2} $&  0.737917&  0.187&  -0.210&    0.005&    0.007\\
&$5f_{5/2} $&$  7d_{5/2} $&  0.553263& -0.036&  -0.052&    0.000&    0.001\\
&$5f_{5/2} $&$  8d_{5/2} $&  0.740531&  0.031&   0.046&    0.000&    0.000\\[0.5pc]
 $\alpha_0^{\rm tail}\textrm{(DF)}$ &&&&&&0.762& 0.762\\
 $\alpha^{\rm core}\textrm{(RPA)}$ &&&&&&7.750& 7.750 \\
  $\delta \alpha_0^{\rm core}\textrm{(RPA)}$&& &&&&-0.050&-0.050 \\
Total &&&&& & 13.523 & 15.073\\
\end{tabular}
\end{ruledtabular}
\end{table*}

\section{ Scalar and tensor polarizabilities  in  the $5f_{5/2}$
ground state of Th$^{3+}$}

We calculate the tensor  polarizability $\alpha_{2}$ of  Th$^{3+}$
in a state $v$ using a sum-over-states approach  \cite{mar-pol-04}
\begin{equation}\label{eq1a}
\alpha _{\mathrm{2}}(v)=\sum_{n}I(v-n)\,.
\end{equation}
Here
\begin{equation}\label{eq2a}
I(v-n)=A(vn)\frac{Z_{vn}^{2}}{E_{n}-E_{v}}\,,
\end{equation}
where
\begin{eqnarray*}
A(nv) &=&-4\sqrt{\frac{5j_{v}(2j_{v}-1)}{6(j_{v}+1)(2j_{v}+1)(2j_{v}+3)}} \\
&&\times (-1)^{j_{v}+j_{n}+1}\left\{
\begin{array}{lll}
j_{v} & 1 & j_{n} \\
1 & j_{v} & 2
\end{array}
\right\}
\end{eqnarray*}
and $Z_{vn}$ is a reduced electric-dipole matrix element.

The calculation of the $\alpha_{2}(5f_{5/2})$  is
divided into three parts:
\begin{eqnarray}\label{main}
\alpha_2^{\mathrm{main}}(5f_{5/2})
&=&\sum_{n=6}^{8}I(5f_{5/2}-nd_{j})+%
\sum_{n=5}^{6}I(5f_{5/2}-ng_{7/2});\   \nonumber  \label{eq5} \\
\delta \alpha_2^{\mathrm{core}}(5f_{5/2})
&=&\sum_{n=3}^{5}I(5f_{5/2}-nd_{j}); \nonumber
\\
\alpha_2^{\mathrm{tail}}(5f_{5/2})
&=&\sum_{n=9}^{50}I(5f_{5/2}-nd_{j})+\sum_{n=7}^{50}I(5f_{5/2}-ng_{7/2})\,.
\end{eqnarray}

 We present the details our calculations of
tensor polarizabilities $\alpha_2$  for the ground
state $5f_{5/2}$ in Table~\ref{tab-comp2}. We use experimental energies from \cite{expt}.
 Electric-dipole matrix elements evaluated in the the
third-order and all-order SD approximations are given in columns
labeled $Z^\text{(DF+2+3)}$ and $Z^\text{(SD)}$. The corresponding
contributions to the tensor polarizability are given in
columns labeled  $\alpha_2^{\rm (DF+2+3)}$ and  $\alpha_2^{\rm (SD)}$.
The contributions $\alpha_2^{\mathrm{tail}}$ and $\delta \alpha_2^{\mathrm{core}}$
are found to be very small and are calculated in the DF approximations.  Our final result
obtained in SD approximation is $\alpha_2(5f_{5/2})$ = -6.2~a$_0$$^3$.

We calculate the  scalar dipole polarizability $\alpha_0$ of Th$^{3+}$
in $5f_{5/2}$ ground state   using the expression
(Refs.~\cite{mar-pol-04}):
\begin{equation}\label{eq6}
\alpha _{{\rm 0}}(5f_{5/2})=\sum_{n}[I_{S}(nd_{3/2})+I_{S}(nd_{5/2})+I_{S}(ng_{7/2})],
\end{equation}
where
\begin{equation}\label{eq7}
I_{S}(nlj)=\frac{1}{9}\frac{Z^2_{5f_{5/2},nlj}}{%
E_{nlj}-E_{5f_{5/2}}}.
\end{equation}
  The breakdown of the contributions to the scalar dipole polarizability together with final result
for the ground state $5f_{5/2}$ in
Th$^{3+}$ are presented in Table~\ref{tab-comp6}. Again, both third-order and
all-order results are listed.  We use the
same designations as in Table~\ref{tab-comp2}.
 We also calculate  the polarizability $\alpha_{\rm
core}$ of the radonlike ionic core in
Th$^{3+}$. Detailed discussion for the $\alpha_{\rm core}$ in Na,
K, Rb, Cs, and Fr atomic systems was presented by Safronova {\it
et al.\/} in Ref.~\cite{safr-alk}.  We evaluate $\alpha_{\rm
core}$ using random-phase approximation (RPA). We find
$\alpha_{\rm core}{\rm (RPA)}$ to be equal to 7.750~(a$_0$)$^3$ in
a.u.. This value was used to obtain our final result for the scalar
ground state polarizability  $\alpha_0(5f_{5/2})=15.1~a^3_0$.
We note that unlike the case of neutral Fr,  the core contribution is
very large, 50\%.
The calculation of the ground state polarizabilities for Fr-like
Th provides another test of the quality of atomic-structure
calculations. There are no experimental results for the Th~IV
polarizabilities at this time. An accurate measurement of the Th~IV
polarizability, combined with these calculations,  may be used to derive the values of the
$5f_{5/2}-6d_{3/2}$ E1 matrix elements and to evaluate the
accuracy of the RPA core value.

\section{Conclusion}
In summary,  a systematic relativistic  MBPT study of energies
 of   $ns_{1/2}$ ($n$= 7-10), $np_j$ ($n$ =
7-9), $nd_j$ ($n$= 6-8),  $nf_{j}$ ($n$= 5-7),  and $ng_{j}$ ($n$=
5-6) states in Fr-like  thorium   is presented. The energies are in good agreement
with existing experimental energy data and provide a theoretical
reference database for line identification. A systematic
 all-order SD study of reduced E1 matrix elements and transition rates for
the 96  electric-dipole transitions in  Th$^{3+}$ is conducted.
Lifetimes are calculated in the SD approximation for $nl_j$ levels.
Multipole  matrix elements ($7s_{1/2}\ - 6d_j$, $7s_{1/2}\ -
5f_j$, and $5f_{5/2}\ - 5f_{7/2}$)  are evaluated to obtain the
lifetimes of the $5f_{7/2}$  and $7s_{1/2}$
states.  Scalar and tensor polarizabilities for the  Th$^{3+}$
ground state
 are calculated using
relativistic third-order and all-order methods. We believe that
our energy, lifetime, and polarizability results  will be
useful in analyzing existing experimental data and in planning
future measurements.

\begin{acknowledgments}
The work of W.R.J.  was supported in part by National Science
Foundation Grant No.\ PHY-04-56828. The work of M.S.S. was
supported in part by National Science Foundation Grant  No.\
PHY-0457078.
  The work of U.I.S. was supported in part by DOE-NNSA/NV
Cooperative Agreement DE-FC52-01NV14050.
\end{acknowledgments}

%\bibliography{thIV}

\end{document}